\documentstyle[prb,aps,preprint,epsf]{revtex}
\tightenlines
\def\sss{\scriptscriptstyle\rm}

\def\kB{k_{\sss B}}
\def\kF{k_{\sss F}}
\def\kl{k_{\sss F \lambda}}

\def\EF{{\vep_{\sss F}}}

\def\Es{{\vep_s}}
\def\boa{{\rm\bf a}}

\def\bk{{\rm\bf k}}
\def\bq{{\rm\bf q}}
\def\br{{\rm\bf r}}
\def\bv{{\rm\bf v}}

\def\vep{\varepsilon}

\def\be{\begin{equation}}
\def\ee{\end{equation}}
\def\ba{\begin{eqnarray}}
\def\ea{\end{eqnarray}}

\def\di{{\rm d}}
\def\uk{{\char47}}

\begin{document}

\draft

\title{Coherent description of electrical and thermal impurity--and--phonon
limited transport in simple metals}

\author{M.\,Ausloos} \address{G.R.A.S.P. and S.U.P.R.A.S,  B5,  Sart Tilman,
B-4000 Li\`ege, Belgium} \author{K.\,Durczewski and J.\,Ulner}
\address{Institute
for Low Temperature and Structure Research, \\Polish Academy of
Sciences, P.O.Box
1410, \\50-950 Wroc{\l}aw~2, Poland} \date{\today}

\maketitle
\begin{abstract}
The electrical resistivity, thermoelectric power and electronic
thermal conductivity of simple (isotropic) metals are studied in a
uniform way. Starting from results of a variational solution of the
Boltzmann equation, a generalized Matthiessen rule is used in order to
superpose the inelastic (or not) electron-phonon and elastic
electron-impurity scattering cross sections ("matrix elements").  The
temperature dependence relative to these processes is given through
simple functions and physical parameters over the usually investigated
range of temperature for each transport coefficient.  The coherence of
such results is emphasized.
\end{abstract}

\vskip 0.5cm PACS{05.60.+w, 17.15.Jf, 44.90.+c}

\vspace{4ex}

\section{Introduction}

To consider the most simple models is sometimes necessary for studying
the charge carrier and thermal transport in solids. Indeed
mathematical approximations used on more sophisticated models may make
less evident, or even completely suppress, certain effects. For the
transport coefficients the latter statement surely concerns the
thermoelectric power and the thermal conductivity. For studying the
electrical resistivity theoretical methods are surely simpler, better
developed and more reliable.

When there is more than a single scattering source the electrical
resistivity is sufficiently accurately treated in metals under the
assumption of Matthiessen rule \cite{Wil,Zim1,Sha} (see exceptions in
Concluding Remarks).  Further assumptions are necessary to find simple
and effective expressions for the thermoelectric power and the thermal
conductivity. The Gorter--Nordheim rule\cite{Bar}, which is usually
applied to treat the thermoelectric power, is based on the assumption
of the Mott(-Jones) formula validity, however, restricted to the
elastic scattering \cite{PRB00} and the adiabatic approximation
\cite{Jon1,Jon}. For the thermal conductivity the Wiedemann--Franz
law is a usual basis on which the separation of the electronic and
lattice contribution to the experimental thermal conductivity is made,
whence for comparing experimental and theoretical data.  The validity
of the Wiedemann--Franz law is also restricted to elastic collisions.
Clearly to resolve the question of elasticity and inelasticity of the
scattering is essential for comparing experimental data and
theoretical laws in particular for the thermoelectric power and the
thermal conductivity.

The most simple inelastic scattering model of the charge carriers by
{\it phonons} describes most features \cite{PRB96} of the
thermoelectric power in normal metals. Consequently, an approach well
accounting for the energy transfer between the charge carriers and the
scattering quasiparticles should be applied beyond that based on the
Mott formula\cite{Wil} .

It is known that the elastic approximation is sufficient for
considering the scattering of the conduction electrons by {\it static
impurities}.  This leads to a constant, temperature independent
residual resistivity, when the potential is appropriately chosen and
the electron gas is degenerate.  The simplest and well founded
potential which leads to the residual resistivity in metals is the
screened Coulomb potential (or Yukawa potential) provided that
impurity concentration is sufficiently low.

Phenomenological considerations \cite{Wil,Par,Zim2} show that without
assuming a behavior for the electrical residual resistivity one 
cannot explain the
low temperature behavior of the {\it electronic} contribution to the
{\it thermal} conductivity.  At moderate temperatures, along with the
scattering by phonons the impurity scattering explains the magnitude
and the basic features of the temperature dependence of the thermal
conductivity of normal nonmagnetic \cite{Zim2} and even magnetic
\cite{Ras} metals, provided that the magnetic moment in the latter
is periodic and well localized. It was shown indeed in
Ref.\onlinecite{Ras}, devoted solely to the thermal conductivity,
$\kappa$, that the magnetic scattering was of equal importance to the
impurity and the phonon scattering.  However, $\kappa$ was neither
compared to the electrical resistivity $\rho$, nor to the
thermoelectric power $S$, for example calculated in the same
temperature interval in Ref.\onlinecite{JMMM1}. Moreover the final
results were not presented through standard functions such as the
Bloch--Gr\"uneisen ones. We shall do so in the present paper, but
confining ourselves to non magnetic impurity and phonon scattering
only.

As in our previous papers we shall apply the Ziman variational
approach \cite{Zim2} within the Boltzmann equation formalism, an
approach in which the notion of "relaxation time" is not used. Thus,
the relationship to the Mott formula is not straightforward \cite{PRB00}.
However, terms responsible for elastic and inelastic scattering can be
easily selected, at least when the conduction band is parabolic.
Within this approach one can generalize the Matthiessen rule: we
consider a superposition of the transition probabilities themselves
within the Boltzmann equation formalism \cite{Sha,JMMM1} instead of
the (integrated) relaxation time components. Therefore the electrical
resistivity, the thermoelectric power and the thermal conductivity are
considered here with respect to impurity and phonon scattering on the
same ground, {\it i.e.} both model and approximations. For this
purpose it is sufficient to introduce a single parameter which
accounts for the mutual contribution of the phonon and the impurity
scattering to the transport coefficients.


A similar model accounting for the scattering by phonons and
impurities was considered in the fifties \cite{Wil} within the Kohler
variational formalism \cite{Wil,Zim2}. However, the mathematical
method was more complicated and less physical than that developed
later by Ziman \cite{Zim2}. It did not allow
\cite{Wil,PRB96,Son} to get qualitatively correct results for the
thermoelectric power in contrast to those for the electrical
resistivity and the electronic thermal conductivity.

The finding of an accurate solution of the (Bloch--)Boltzmann
equation is not a past problem of the fifties or the seventies but still a
present--day problem. Numerical methods are used for this purpose.
Recently \cite{Sch}  the Allen method \cite{All} of solving the
equation with a good numerical convergence has been used to study the
influence of the electron--electron scattering on the electrical and
thermal conductivity. A comparative discussion of the approximation
made while using the Ziman variation method \cite{Zim2} and the Allen
one \cite{All} is given in Sec.V.


In the following Section II we shortly describe the Ziman variational
method and the general results when applied to the model here defined.
Certain general results and formulae are also given in Appendix A for
completeness.  In Section III the matrix elements describing the
phonon and impurity scattering are calculated and final formulas for
the electrical resistivity, electronic thermal conductivity and
thermoelectric power are systematically presented.  In Section IV the
temperature dependence of the transport coefficients and that of the
Lorenz number is numerically illustrated. The visual presentation is
made by using dimensionless functions. An analysis of appropriate
material constants allowing to estimate the magnitude of the final
transport coefficients is found in Appendix B. Section V contains the
concluding remarks.

\section{Model, method and approximations.}

We assume that the current carriers are free electrons with a
parabolic isotropic energy band $\vep(\bk)=\hbar^2 \bk / 2m$ spectrum.
We also assume that the concentration of the electron gas (of both
subbands with spin up or down) $n_e =
(1/3\pi^2)(2m/\hbar^2)^{3/2}\EF^{3/2}$ (or the Fermi energy $\EF$) is
sufficiently high to consider the gas to be degenerate. Then one can
assume that the chemical potential is $\zeta(T) \cong \EF$, and all
relevant functions of $z = \zeta /\kB T$ can be considered as
asymptotic expressions of the argument $\EF /\kB T$ . In particular
the Fermi-Dirac integrals

\be \label{1}
{F}_n(z)=\int_0^{\infty}\!\di x\,\frac{x^n}{1+\exp[x-z]},
\mbox{\qquad}{\rm \, with} \mbox{\qquad}  z=\frac{\zeta(T)}{\kB T}
\ee
\noindent
can be represented as

\be \label{2}
{F}_n(z) \cong z^{n+1}/(n+1), \mbox{\qquad}{\rm \,
with} \mbox{\qquad}  z \cong \EF/ {\kB T}
\ee
\noindent
for sufficiently large $z$.

The carriers are assumed to be scattered by static, nonmagnetic
impurities, which concentration is sufficiently low to have a linear
influence on the electrical resistivity . The scattering of carriers
by phonons will be considered on the same footing. However, the phonon
(quasiparticle) scattering will be treated as inelastic while the
impurity scattering is elastic in its nature, since the impurity is
usually assumed to be more heavy than the electron effective mass.
Even though some calculations can be performed without assuming the
form of the potential) we assume the form of the impurity potential to
be the screened (Yukawa) potential

\be \label{3}
U(\br) = - \frac{Ze^2 \exp(-\lambda r)}{r}
\ee
\noindent
with $r =|\br|$, $Z = Z_{impurity} - Z_{host\,ion}$ the effective
valence and the screening constant $\lambda$. The Fourier transform
$U(\bq) = 4\pi Z e^2/(\lambda^2 + \bq^2)$ of the above function is
responsible for the magnitude of the carrier scattering; $\bq$ with $q
\equiv |\bq|$ is the electron wave vector transferred during the
scattering.  The transport relaxation time following from $U(\bq)$ --
and consequently the residual resistivity -- is temperature
independent for a degenerate electron gas.  Their magnitude is related
to the Fermi wave vector $\kF$ or the energy $\EF$ through the function
\cite{Zim1}

\be \label{4}
2 \int_0^{2\kF}\frac{\!\di q\,q^3}{(\lambda^2+q^2)^2} =
\ln(1+4\kl^2) - 4\kl^2/(1+4\kl^2) \equiv \mit\Phi (2\kl) ,
\ee
\noindent
where $\kl = \kF/\lambda$ .

Note that the standard considerations \cite{Zim1} yield that
$\lambda^2 \sim {n_e}^{1/3}$ and thus one can estimate $2\kF =
(3\pi)^{1/6}(a_H\kF)^{1/2} \cong 1$ since the effective Bohr radius
$a_H = \hbar/(me^2)$ is of the order of $\kF$.

The phonon system and the electron--phonon scattering are considered
in the Bloch approximation, {\it i.e.} phonons in the harmonic
approximation and the electron--phonon interaction in the deformation
potential approximation.  We restrict the magnitude of the phonon
wave vector, $q \le q_D$, to be within the Debye sphere. Then, for
sufficiently high values of the Fermi energy, $z =
\rightarrow \infty$ (see,e.g., Appendix B in Ref.\onlinecite{PRB96}),
the phonon limited transport coefficients are expressed in terms of the
Bloch(--Gr\"uneisen) functions

\be \label{5}
{\cal F}_n(t)=\int\limits_0^{1/\uk t} \,\frac{dx\,x^n}{({\rm
e}^x-1)(1-{\rm e}^{-x})}\:.
\ee


A relation between the deformation potential approximation and the
method based on the dynamical structure factor introduced by van Hove
\cite{Hov} is described in Sec. V.


As we mentioned in the Introduction we apply the Ziman variation
method \cite{Zim2} to consider the combined effect of the impurity and
phonon scattering on the transport coefficients. The method consists
in calculating the elements of a matrix describing the scattering and
two vectors governing the flow of the electric charge and heat.  The
''exact'' results correspond to an infinite order matrix and infinite
dimension current vectors (see, {\it e.g.} Appendix A and Sec. IV in
Ref.\onlinecite{PRB00}). Here, we truncate the matrix and vectors to
four elements and two components, respectively. This is the lowest
order approximation, which accounts well for both the electrical and
thermal current if the variational trial functions are appropriately
chosen\cite{Zim2,ZPh91}. This choice is known to be

\be \label{6}
\mit\Omega_i(\bk)=({\bv\cdot\boa})\,(\vep-\zeta)^{i-1}\:,
\ee
where $i=1$ and $2$,  while $\boa$ denotes the unit vector in the direction of
the applied temperature gradient or that of the external electric field causing
the charge flow, and $\bv =(\hbar/m) \nabla_k \vep $ is the electron velocity.

In this approximation the transport coefficients : the electrical
resistivity $\rho$, the thermoelectric power or the Seebeck
coefficient $S$ and the (electronic) thermal conductivity $\kappa$ for
cubic crystals, are given as such

\be \label{7}
\rho=(P_{11}P_{22} - P_{12}^2)\,\mit\Pi^{-1}\:,
\ee

\be \label{8}
S=[P_{22}J_1U_2-P_{12}(J_1U_2+J_2U_1)+P_{11}J_2U_2]\,(T \mit\Pi)^{-1}\,
\ee

\be \label{9}
\kappa=(J_1^2U_2^2+J_2^2U_1^2 -2 J_1J_2U_1U_2)\,(T \mit\Pi)^{-1}\:.
\ee
with

\be
\mit\Pi = P_{22}\,J_1^2-2P_{12}J_1J_2+P_{11}J_2^2,
\ee
\noindent
the elements of the scattering matrix being

\ba \label{10}
P_{ij}&=&\int\!\di\bk\,\int\!\di\bk'\,C(\bk,\bk')\,u_{ij}(\bk,\bk')\:,
\nonumber\\ u_{ij}(\bk,\bk')&=&[{\mit\Omega}_i(\bk)-{\mit\Omega}_i(\bk')]\,[
{\mit\Omega}_j(\bk)-{\mit\Omega}_j(\bk')]\:,
\ea
\noindent
and the components of the trial currents \footnote{The factor ({\bf
v.a}) in the expression for the trial currents in
Ref. 5 was inadvertently omitted. In the trial functions (12)-(13)
this factor is used as in Ref. 2, and in our previous papers where it
was denoted ({\bf k.a}).}

\be \label{11a}
J_i=-e\int\!\di\bk\,\left(-\frac{\di f}{\di\vep}\right)\,
({\bv\cdot\boa}){\mit\Omega}_i(\bk)~,
\ee
\be \label{11b}
U_i=\int\!\di\bk\,\left(-\frac{\di f}{\di\vep}\right)\,
({\bv\cdot\boa})(\vep-\zeta){\mit\Omega}_i(\bk)~,
\ee
where $C(\bk,\bk')$ is the transition probability to be determined on
the ground of an appropriate Hamiltonian describing the scattering.
The function $f$ denotes the Fermi--Dirac one,
$f=1/[1+\exp(\vep-\zeta)/\kB T]$, and $-e (e>0)$ is the electron
charge.

The integrals in (\ref{10}) and (\ref{11a})-(\ref{11b}) should be
taken over the Brillouin zone but in our free electron considerations
we let

\be \label{12}
\int\!\di\bk \ldots \equiv \frac{2}{(2 \pi)^3}
\int\limits_{0}^{\infty} {\rm d}{\vep} \int\!\left(\frac{\di
S_{\vep}}{|\nabla_k\vep|}\right)\ldots\:,
\ee
where $\di S_{\vep}/{|\nabla_k\vep|}$ is an element of the isoenergy
surface.  The integrals with respect to the electron energy in (10)
for the degenerate electron gas are described in Appendix B of Ref. 8.
In the case of elastic scattering by impurities, as here, the
integrals can be treated as simply. Namely, we take advantage of the
approximate equalities

\be \label{30}
\int\limits_0^{\infty}\,\di \vep (-\di f/\di\vep)(\vep - \zeta)
g(\vep) = (\pi^2/3) (\kB T)^2 [\di g/\di \vep]_{\zeta} \:,
\ee

\be \label{29}
\int\limits_0^{\infty}\,\di \vep (-\di f/\di \vep)(\vep - \zeta)^2
g(\vep) = (\pi^2/3) (\kB T)^2 g(\zeta) \:.
\ee
where $g(\vep)$ is an arbitrary function.

The trial currents (\ref{11a})-(\ref{11b}) can be represented
\cite{PRB96} as acombination of the Fermi--Dirac integrals
(see Appendix B) : $J_1 = -J_0(\kB T)^{3/2}L_0(z)$, $J_2 = -J_0(\kB
T)^{5/2}L_1(z)$, $U_1 = -U_0(\kB T)^{5/2}L_1(z)$, $U_2 = -U_0(\kB
T)^{7/2}L_2(z)$, where $U_0$ and $J_0 = eU_0$ are material constants.

For the electron--phonon system the $P$--matrix elements can be
derived (see Ref.\onlinecite{PRB96}) without any further
approximation.  They are expressed by double integrals with respect
to the electron energy and the transferred electron momentum
(wave vector). For a Debye cutoff and in the limit $z
\rightarrow \infty$ they can be represented\footnote{For further
convenience, the factor $t^5$ used in Ref.8 is introduced here
directly in the definition of ${\cal P}_{ij}$.} by $P_{ij} = P_0 (\kB
T)^{i+j-2} {\cal P}_{ij}$, where $P_0$ is a material constant and
${\cal P}_{ij} = {\cal P}_{ij}(t)$ are dimensionless combinations of
the the Bloch--Gr\"uneisen functions of the argument $t = T/T_D$
($T_D$ is the Debye temperature).

For the impurity scattering an expression similar to the above is
readily obtained. In this case the ${\cal P}_{ij}$ 's are simply
related to $\mit\Phi (2\kF/\lambda)$ defined in (4). The explicit form
of the functions ${\cal P}_{ij}$ will be given in the next Section.
The magnitude of $P_0$ and that of other material constants are estimated in
Appendix B.

In view of the above, by calculating the scattering matrix elements in
the degenerate limit, retaining only ${\cal P}_{11} (\gg {\cal
P}_{12}^2 / {\cal P}_{22})$ in case of the electrical resistivity and
the lowest order terms in the $1/z$ in the trial currents (12)-(13),
we obtain

\be \label{13}
\rho=\frac{P_{11}} {J_1^2} =  \frac{P_0 {\cal P}_{11}}{J_0^2 \EF
^3}\:,
\ee

\be \label{14}
S = S_0(t) + S_1(t) \frac{\kB T}{\EF}\:,
\ee

\be \label{15}
\frac{1}{\kappa} = \frac{P_{22}} {(\pi^2 /3){\cal L}_0 T (\kB T)^3
J_1^2} =  \frac {P_0}{U_0^2}\, \frac {T {\cal P}_{22}}{(\pi^2/3)^2
(\kB T)^2 \EF^3}\: ,
\ee
where $J_1$ in (\ref{13}) and (\ref{15}) is the trial current in its
asymptotic form and ${\cal L}_0 = (\pi^2 /3)(\kB /e)^2$ is the
Sommerfeld value of the Lorenz number.
The explicit form of the functions $S_0(t)$ and $S_1(t)$ of (\ref{14})
-- having a more complicated dependence on the matrix elements ${\cal
P}_{ij}$ -- are discussed in Section III B.

\section{Scattering matrix contributions.}

As seen from (\ref{13})-(\ref{15}) it remains to calculate the
elements of the $P$--matrix in order to find the final expressions for
the transport coefficients. In our case the crux is to take into
account the combined effect of the scattering by impurities and
phonons.

As we mentioned in the Introduction we apply the Matthiessen rule in a
more generalized way \cite{Sha,Ras,JMMM1}, i.e. to the microscopic
transition probabilities C(\bk,\bk') themselves appearing in
Eq.(\ref{10}) rather than to the final coefficients: the electrical
and thermal resistivity.  The variational method enables us to use
the same approximation to both these coefficients and the
thermoelectric power through the integrated probabilities, {\it i.e},
we assume

\be \label{16}
C(\bk,\bk') = C^{(imp)}(\bk,\bk') + C^{(ph)}(\bk,\bk')\,.
\ee
\noindent
This means that the resultant scattering probability for the phonon
and impurity scattering can be simply considered as a sum of the
scattering probabilities of the impurities (the phonon scattering
being quenched) and by phonons (those of the pure crystal). This, of
course, implies that the elements of the resultant scattering matrix,
(\ref{10}), are also the sums of the corresponding matrix elements for
the phonon and impurity scattering. Therefore, we write

\be \label{17}
P_{ij} = P_{ij}^{(imp)} + P_{ij}^{(ph)}\:.
\ee
\noindent
in formulae (\ref{7}) - (\ref{9}).

The electron-phonon contribution $P_{ij}^{(ph)}$'s have been derived
in previous papers, {\it e.g.}, in Ref. \onlinecite{PRB96} they are
expressed in terms of the Bloch(-Gr\"uneisen) functions (\ref{5}).
The appropriate expressions for the electron-impurity contribution
$P_{ij}^{(imp)}$ following from the Yukawa potential (\ref{3}) are
given in this Section, but first we write down the final expression
for the electronic thermal conductivity with the influence of the
residual electrical resistivity as a parameter. For this purpose we
have to separate elastic and inelastic contribution to $P_{22}$ and
therefore we write explicitly the matrices $u_{ij}$ of the kernels
in (10).  For the parabolic band $\vep(\bk)=\hbar^2\bk / 2m$ they can
be represented as

\ba \label{18}
u_{11}&=&(\hbar/m)^2\,(\bq\cdot\boa)^2\,,\\
u_{12}=u_{21}&=&(\hbar/m)^2\,[(\vep-\zeta)+(\vep'-\vep)]\,(\bq\cdot\boa)^2\,
,\nonumber\\
u_{22}&=&(\hbar/m)^2\,[(\vep-\zeta)^2+2(\vep'-\vep)(\vep-\zeta)]
(\bq\cdot\boa)^2+(\vep'-\vep)^2[(\bq\cdot\boa)^2+(\bk\cdot\boa)^2]\,.\nonumber
\ea
where ($\vep-\vep$) is the transferred energy during the scattering,
$\bq=\bk'-\bk$ is the transferred wave vector and terms linear in
$\bk$ and $\bk'$, which do not contribute to the integrals, have been
neglected. It is seen that the second component in $u_{12}$ and the
second and third one in $u_{22}$ contribute only when a change of the
electron energy in the scattering is taken into account. It means
that the latter terms describe the effects of {\it 
inelasticity}\footnote{These terms describe the {\it basic} inelastic
contribution.  For the phonon scattering the single term of $u_{11}$
and the first terms in the case of $u_{12}$ and $u_{22}$ are
expressed by the Debye functions in case of pure elastic scattering
and by the Bloch(-Gr\"uneisen) functions in case of inelastic
scattering (see Ref.8). The terms proportional to $(\vep'-\vep)$ and
$(\vep'-\vep)^2$ in $u_{12}$ and $u_{22}$ lead to an essential
qualitative effect in the thermoelectric power and the thermal
conductivity.} in the scattering. With the definition

\be \label{19}
P_{11}^{(ph)} = P_0^{(ph)}\,{{\cal
P}}_{11}^{(ph)}(t)\,,\,\,\,\,\,\,\,  {\rm with}\, t = T/T_D
\ee
and where $P_0^{(ph)}$ is a material constant (see Appendix B), we get

\be \label{20} P_{22}^{(ph)} = P_0^{(ph)}\,(\kB T)^2\, [{{\cal
P}}_{22}^{(1,ph)}(t) + {{\cal P}}_{22}^{(2,ph)}(t)] \equiv  P_{22}^{(1,ph)} +
P_{22}^{(2,ph)}\,, \ee where

\ba \label{21}
{{\cal P}}_{22}^{(1,ph)}(t)&=&\frac{\pi^2}{3}\,{{\cal
P}}_{11}^{(ph)}(t) \:,\nonumber\\
{{\cal P}}_{22}^{(2,ph)}(t)&=&t^5\,[\frac{1}{3}\,{\cal F}_7(t)+
\frac{\Es\EF}{(\kB T)^2}\,{\cal F}_5(t)-\frac{\Es}{2\kB T}\,{\cal
F}_6(t) ]~\:,
\ea
and $\Es$ is a constant (see Appendix B).

The explicit form of ${{P}}_{11}^{(ph)}(t)$ describing the electrical
resistivity is

\be \label{22}
{{\cal P}}_{11}^{(ph)}(t) = t^5\,{\cal F}_5(t)~\:.
\ee
The two terms in $u_{12}$, accounting respectively for the elastic and
inelastic contributions, after integration with the transition
probability for the phonon scattering can be represented by a single
term,

\be \label{23}
P_{12}^{(ph)} = P_{21}^{(ph)} = P_0^{(ph)} (\kB T) {{\cal
P}}_{12}^{(ph)}(t)
\ee
with

\be \label{24}
{{\cal P}}_{12}^{(ph)}(t) = {{\cal P}}_{21}^{(ph)}(t) =
\frac{1}{2}\,t^5\,{\cal F}_6(t)~\:.
\ee

Two material constants influence the magnitude of ${{\cal
P}}_{22}^{(2,ph)}(t)$.  These are the Fermi energy, $\EF$, and
$\Es=2mv_s^2$, where $v_s$ is the sound velocity (see the Table I in
Appendix).  The latter energy (or more precisely $\Es /4$) is a
threshold energy for the phonon emission in the electron--phonon
scattering process (see Appendix J in Ref.\onlinecite{Kit}).

Assume now that the charged (static) impurity concentration is so low
that one can consider them to be statistically independent. Then the
scattering of the conduction electrons by a single impurity
contributes solely to the $P_{ij}$ matrix elements and they are linear
in their concentrations. In consequence the electrical resistivity is
linear in the electron concentration and for the degenerate electron
gas it is also temperature independent.

The latter statement can be easily confirmed by the calculation of
$P_{11}^{(imp)}$ in the metallic limit. For further convenience we
assume $P_{11}^{(imp)}/P_{0}^{(ph)} \equiv \mit\Gamma$ and due to the
Sommerfeld expansion of terms of (\ref{18}) corresponding to an
elastic scattering we have $P_{22}^{(imp)} = P_{22}^{(imp,1)} =
P_{0}^{(ph)}\, (\pi^2/3)(\kB T)^2 \mit\Gamma$. Therefore, the final
expressions for $P_{11}$ and $P_{22}$ describing the impurity and
phonon scattering are

\ba \label{25}
P_{11}= P_0^{(ph)}\,{{\cal P}}_{11}(t),\, \:\mbox{\qquad} {\rm with}
\mbox{\qquad\qquad}
{{\cal P}}_{11}(t) = P_0^{(ph)}\,[\mit\Gamma + {{\cal
P}}_{11}^{(ph)}(t)] \nonumber \\
P_{22} = P_0^{(ph)}(\kB T)^2\,{{\cal P}}_{22}(t),\,\:\mbox{\qquad}{\rm
with} \mbox{\qquad}
{{\cal P}}_{22}(t) = \frac{\pi^2}{3}[\mit\Gamma + {{\cal 
P}}_{11}^{(ph)}(t)] + {{\cal
P}}_{22}^{(2,ph)}(t)\,\, .
\ea

\subsection{Electrical resistivity and thermal conductivity in metals}

The above expressions for the matrix elements of the $P$--matrix yield the
electrical resistivity

\be \label{26}
\rho = \rho_{0,ph}\,[\mit\Gamma + {{\cal P}}_{11}^{(ph)}(t)]\, ,
\ee
where $\rho_{0,ph}$ is a constant (see Appendix B), and the electronic
thermal conductivity

\be \label{27}
\kappa = \kappa_{0,ph} \,\frac{t}{{{\cal P}}_{22}(t)}\,,
\mbox{\qquad\qquad}
\kappa_{0,ph} = \frac{\pi^2}{3} \frac{{\cal L}_0 T_D}{\rho_
{0,ph}}\,.
\ee
Note that the Lorenz number reads

\be \label{28}
{\cal L} = \frac{\rho \, \kappa}{T} = {\cal L}_0 \frac{\pi^2}{3}
\frac{{{\cal P}}_{11}(t)}{{{\cal P}}_{22}(t)}
\ee
and reduces to the Sommerfeld value ${\cal L}_0$ for elastic
scattering, when one neglects ${{\cal P}}_{22}^{(2,ph)}$ in (25).

The results (26) describing the electrical resistivity are obviously
standard ones and do not require any comments. As concerns the
electronic thermal conductivity and the Lorenz number similar
expressions in terms of the Bloch--Gr\"uneisen functions as those in
the present paper were obtained by Wilson \cite{Wil} and were
numerically examined in Ref. \onlinecite{Par} manifesting the
influence of the impurity scattering.  The Wilson expression for the
thermal resistivity in our notation reads

\be \label{33}
\frac{2\pi^3}{3} \,t^5\,{\cal F}_5(t)+\frac{8\pi^2\Es\EF}
{(\kB T)^2}\,t^5\, {\cal F}_5(t)-\frac{1}{3}\,t^5{\cal F}_7(t) ~\:.
\ee

\subsection{Seebeck coefficient of metals.}

In order to discuss the thermoelectric power, $S$, we introduce, as
for the discussion of the thermal conductivity in
Ref.\onlinecite{Ras}, a relative order of magnitude parameter $\varphi
= P_0^{(imp)} / P_0^{(ph)}$, in terms of which for the Yukawa
potential (\ref{3}) we have $\mit\Gamma (\EF) = \varphi \mit\Phi (2\kF
/\lambda)$. It will be also convenient to introduce here the function
$\mit\Psi (x) = x^4/(1+x^2)^2$ in terms of which we express $[\di
\mit\Phi /\di \vep]_\EF = (\EF)^{-1}\mit\Psi (2\kF /\lambda)$ .
In so doing the off-diagonal component of the $P$--matrix responsible
for the impurity scattering is linear in $\kB T/\EF$ and reads

\be \label{31}
P_{12}^{(imp)}(T) = P_0^{(imp)}\,(\kB T)\,{{\cal
P}}_{12}^{(imp)}\,, \mbox{\qquad\qquad} {{\cal P}}_{12}^{(imp)} =
\frac{\pi^2}{3}\,\varphi\,\mit\Psi(2\kF/\lambda)\,\frac{\kB
T}{\EF}~\:.
\ee
When these expressions are inserted into (\ref{8}) and linear terms in $\kB T
/\EF$ are only accounted for, one obtains the following expressions for the
coefficients in (\ref{14})

\ba \label{32}
S_0(t) = \frac{\pi^2}{3}\,\frac{\kB}{e}\, \frac{{{\cal
P}}_{12}^{(ph)}(t)}{{{\cal P}}_{22}(t)}\,,\mbox{\qquad}
S_1(t) = -\frac{\kB}{e}\left\{\frac{\pi^2}{2}\, \left[1 +
\frac{\pi^3}{3}\,\frac{{\cal{P}}_{11}(t)}{{{\cal P}}_{22}(t)}\right] -
(\frac{\pi^3}{3})^2\, \frac{\varphi \mit\Psi (2\kF /\lambda)}{{{\cal
P}}_{22}(t)}\right\}\,.
\ea

Note that the impurity scattering has an influence on the term
$S_0(t)$ only through the matrix element ${{\cal P}}_{22}(t)$ in the
denominator.  If $\varphi = 0$, the above formula corresponds to that
obtained in Ref.\onlinecite{PRB96} for pure phonon limited
thermoelectric power.

Notice that the pure phonon limited thermoelectric power obtained
within the Wilson formalism\cite{Son} do not represent any
characteristic features of the experimentally observed $S$ behavior
in simple monovalent metals.  Such results for $S$ obtained in
Ref.\onlinecite{PRB96} were already discussed there in terms of
various approximations.

\section{Temperature dependence of the transport coefficients and the Lorenz
number.}

To compute the transport coefficients as a function of $t=T/T_D$ we
assume the value of $2\kF/\lambda = 0.75$ and $ 0 \le \varphi \le 1 $.
The temperature dependence of the reduced electrical resistivity
following from (26), $\rho/\rho_{0,ph}$, which is shown in Fig. 1, and
the Seebeck coefficient in Fig. 2 have been computed for $\varphi =
0.0, 0.2, 0.5, 1.0, 2.0$ and the corresponding values of $\mit\Gamma =
\varphi\mit\Phi$(0.75) = 0.0, 0.0173, 0.0259, 0.0431,
0.0862, 0.173 respectively. As it is seen the impurity scattering
slightly lowers the maximum of $S$ following from the transfer of the
energy between the electron and phonon system\cite{PRB96}. The values
of $\Es$ and $\EF$ have been chosen to be the same as those in
Ref.\onlinecite{PRB96}, namely $\Es=6.0$\,K and $\EF=1.5$\,eV. The
electronic thermal conductivity is considerably affected by the
impurity scattering, what is seen from the data represented in Fig. 3
and Fig.4.  In Fig.3 the reduced thermal conductivity ($\kappa
/\kappa_{0,ph}$) is shown as a function of $t$ in the low temperature
range such as in Fig. 4.16 of Ref.\onlinecite{Par}, where the results
of Wilson's considerations\cite{Wil} were already illustrated. In
Fig. 4 the results are shown in the same temperature range as the
thermoelectric power in Fig. 2. We note that he dependence of the
thermal conductivity as a function of temperature here obtained by the
Ziman variational method is qualitatively the same as that following
from Wilson's considerations\cite{Wil,Par}. Furthermore, at
physically realistic temperatures the values of $\kappa$ obtained by
the latter method approach a saturation, which is characteristic of
monovalent metals. Under the assumption of the same values of the
parameters describing the impurity scattering the saturation of
$\kappa$ following from Wilson formula occurs at a so high temperature
that it ceases to be meaningful. According to Ziman\cite{Zim3}, who
examined the Wilson formula, physically reasonable results for
monovalent metals (saturation and lack of minimum at intermediate
temperatures) are obtained when the Umklapp scattering is taken into
account. Experimental results for rare earth intermetallic compounds
exhibit the existence of a minimum at temperatures such as those seen
in Figs. 3 and 4 (see e.g. Fig. 7 in Ref.\onlinecite{Ras2} and references
therein), in which the magnetic scattering are also taken into
account.

The reduced Lorenz number, $\cal L /\cal L$$_0$, as a function of $t$
is shown in Fig. 5 for the same values of the parameters as $\kappa
/\kappa_{0,ph}$ in Fig. 4.  The behavior of this quantity as a
function of $t$ is qualitatively the same as that following from the
Wilson considerations \cite{Wil,Par}. Note that for only phonon
scattering $\cal L$(0) = 0, whereas $\kappa (0)$ tends to infinity at
zero temperature.

\section{Concluding remarks.}


Using a simple physical electron model, applying the Ziman
variational method within the Boltzmann equation approach and using
the same level of approximations for each property we have calculated 
the phonon-limited
electrical resistivity, thermoelectric power and the electronic
thermal conductivity of a simple metal and studied how the presence
of charged impurities influences these transport coefficients.  The
method allows us to determine the temperature dependence of the
coefficients in a wide temperature interval and to express them in
terms of simple and standard functions already useful for examining 
experimental data.
However we are aware that to take into account a more
complicated band structure is surely necessary in many cases. In
this purpose the Allen method\cite{All} is surely more appropriate though the
task needs to bear upon numerical technicalities from the very beginning. Note
that the latter method is strictly applicable to metallic systems, when the
electron gas is degenerate, while by using the method of this present
paper one advantage is that one can easily consider semiconductor or 
semimetallic systems as well
\cite{PRB00}.

A comment is in order on the results of the present paper obtained in 
the limit of the
degenerate electron gas.  As concerns the electrical resistivity and
the electronic thermal conductivity (or resistivity) the final
results are qualitatively the same as those obtained previously
in Refs.\onlinecite{Wil,Par,Zim2}. In the case of the thermoelectric 
power however we
can obtain a new evidence as a conclusion,
i.e. the maximum at intermediate temperatures is caused by the
energy transfer between the conduction electrons and
phonons\cite{PRB96} and is lowered as a result of the elastic scattering
of the electrons on impurities.


The final formal expression for the electronic thermal resistivity
reads as a simple expression: this quantity is proportional to the
scattering matrix element $P_{22}$, much like the electrical
resistivity is proportional to $P_{11}$. The result is in fact
model independent and is a consequence of confining the
scattering matrix to four elements (see Appendix A).


The range of applicability of the method we have used here above cannot be
finally discussed without referring to papers using more general and
sophisticated methods and/or models. In the first place this concerns
the papers in which the electron transport was studied within the
formalism of the Boltzmann equation\cite{All,Sch}. In the second
place a relation to the linear response theory \cite{Jon1,Vil,Ono}
and the quantum transport equation \cite{Liv,Rei1,Rei2} will be
given.  However, by using these methods one cannot obtain simple,
compact, and useful expressions for the three transport coefficients, moreover
being valid in the wide temperature region experimentally studied.
Further complications have also to be expected when taking into account
the phonon drag \cite{Bel}, the virtual recoil \cite{Bla,Nie} or the
electron--phonon renormalisation \cite{Ops,Kai}.

Let us start our discussion with a comparison of our model and
results with the other ones obtained by the Ziman variational method
within the formalism of the Boltzmann equation. As we stressed at the
beginning we confined ourselves to a model of the parabolic electron
band and the Debye model of the phonon system. This oversimplified
model allowed us to gain a coherent and qualitatively correct
description of the three transport coefficients: the electrical
resistivity, thermoelectric power and the thermal conductivity
(resistivity) for a simple and standard model of the scattering.  Any
generalization of these results to an arbitrary band structure
requires numerical studies at an initial stage.
A compact formula in this general case of the Ziman variational 
method has been obtained
only for the electrical resistivity among the three
coefficients, a formula which has been called ''Ziman's
resistivity formula'' \cite{Gri}. It has been used to compute the electrical
resistivity for a realistic Fermi surface.

Again a more exact method is that of Allen\cite{All} though a 
numerical one.  The electrical resistivity is
again simplest to compute but the results for the thermal
conductivity \cite{Sch} and the thermoelectric power \cite{Pic} were
also obtained by this method. The essence of the method is the
separation of the $\bk$-variation during the scattering into two types:
(i) an ''angular variation'' over the Fermi surface, and (ii) a ''radial''
variation with $\vep(k)$ perpendicular to the Fermi surface describing
the inelastic electron scattering. It is obvious from our results for the
thermoelectric power \cite{PRB96} that taking into account inelastic
electron scattering, which is equivalent to considering a sufficient
number of harmonics describing the radial variation within the Allen
formalism, is of major importance for obtaining a correspondence
to experimental results if the scattering of the electrons by
excitations such as phonons is to be accounted. In Ref. \onlinecite{Pic},
in which the lowest order radial harmonic was considered and the
analysis was in fact restricted to the Mott formula, a poor
correspondence to the experimental results for the thermoelectric
power was obtained. Inelastic scattering and the radial harmonics are
also important to describe properly the thermal conductivity. In Ref.
\onlinecite{Sch} the radial harmonics were applied to discuss the
influence of the electron--electron scattering on the thermal
conductivity. According to our knowledge no such considerations were
made to analyze the influence of the electron--phonon scattering
neither on the thermal conductivity nor the thermoelectric power.


The assumed statistical independence between the impurities as
scattering centers restrict our considerations to low impurity
concentration. This is consistent with the experimental fact that the
Matthiessen rule is valid below a critical residual resistivity
\cite{Sha} above which the $T^5$ low temperature dependence
(following, for instance, from the Bloch--Gr\"uneisen dependence of the
electrical resistivity) ceases to be valid. One can hardly estimate such a
critical concentration without analyzing a specific system and experimental
results.

To estimate the magnitude of the transport coefficients out of the
reduced values illustrated on each figure in Sec. IV we present the
order of magnitude of the relevant material constants in Appendix B.
The magnitude of $\Es$ in Table I is the same as in
Ref.\onlinecite{PRB96} and corresponds to the magnitude of the sound
velocity and the effective electron mass(es) of real materials.

We start our discussion of more sophisticated methods than the
Boltzmann equation formalism by noting that in the linear response
theory the validity of the Matthiesen rule, which was assumed by us
in a generalized version, is related to the fact that the self-energy
of the conduction electrons is the same as the sum of the phonon and
impurity self-energy \cite{Kre}. This holds only in the lowest Born
approximation. When higher order processes are taken into account,
{\it i.e.}, when interference between the electron-phonon and
electron-impurity interaction is to be considered, the Keldysh
quantum transport equation formalism \cite{Liv,Rei1,Rei2,Bel,Kel} is
apparently better.

In Ref.\onlinecite{Liv} the same model as in the present paper was
considered to study the influence of the electron-phonon-impurity
interference on the thermoelectric power. A considerable nonlinearity
of $S$ as a function of temperature was found, the stronger the
higher impurity effective charge.  The effect can be hardly compared
with that studied in the present paper since it has been obtained
under an assumption of elasticity for the electron-phonon scattering
whereas the assumption of inelasticity of the electron-phonon
scattering is essential \cite{PRB96} for obtaining the maximum of $S$
found in the present paper and Ref. 8.  Note that as a result of a
weak elastic impurity scattering (considered within the Born
approximation) the bump in $S$ is found to be lowered with the
strength of the impurity interaction or the impurity concentration.

The electron-phonon renormalization \cite{Jon,Kai} leads also to a
nonlinearity of $S$ as a function of $T$.  The effect was studied
under the assumption of elastic electron-phonon interaction or in the
adiabatic approximation. Again there can be no discussion between our
results and those in Ref.\onlinecite{Jon} and \onlinecite{Kai} except 
for the region
of very low temperatures, where the effect of elasticity and
renormalization are not effective and where $S$ is a linear function
of $T$ with the same slope \cite{Jon} as in the pure crystal
\cite{PRB96,Kle}.

Though the impurity scattering is known to play an important role in
the thermal conductivity at least in the low temperature region there
is a very limited number of papers devoted to this coefficient aspect.
The basic difficulties and problems related to applying the methods of
the linear response formalism to treating the electronic thermal
conductivity problem in the pure electron-phonon system are described
in Ref.\onlinecite{Ono}. However, such considerations as those in
Ref.\onlinecite{Ono} have not be generalized for a system with
impurities.  On the other hand the considerations of
Refs.\onlinecite{Jon,Vil}, devoted to thermoelectric effects for a
model of Debye phonons with fixed impurities, have not been used by
these authors in an appropriate way to consider the electronic thermal
conductivity. Therefore, a complete theory of the electronic thermal
conductivity still requires a thorough consideration of the mutual
role of the electron-phonon-impurity interference, phonon
renormalization and elasticity of the scattering, much the same as on
the thermoelectric power.  For comparison to data, in not too good
metals, the phonon heat contribution should be also taken into account
for the thermal conductivity.


A question may arise if our results can be compared to experimental
ones in a straightforward, quantitative way. The answer seems that
these are the alkali metals, which exhibit essentially spherical
Fermi surface to within a few per cent, that should be first of all
considered as materials for the comparison.  It occurs, however, that
the phonon system and the electron--phonon interaction in these
metals are too complicated to be well accounted by the applied
approximations.  It concerns in the first place, (i), the continuous
medium approximation used to treat the phonon system and the
deformation potential approximation applied to the electron--phonon
interaction (see, e.g., Ref.\onlinecite{JMMM1}). In the second place,
(ii), the Debye approximation is insufficient for the alkali metals.
Finally, in the description of the electron scattering we also
neglect the Umklapp processes, which occur important even at very low
temperatures \cite{Gri,Vuc} in the alkali metals.

The approximations (i) mean that the longitudinal phonons of long
wavelength are only considered, whereas the transverse phonons
\cite{Dar} and those of short wavelength (contributing also to the Umklapp
process) have to be taken into account \cite{Dar,Gre}. Due to a
strong elastic anisotropy the Debye approximation (ii) is not
appropriate even at low temperatures since a constant frequency
surface is distorted from the Debye spherical form \cite{Woo,Cow}.
Furthermore, the strong influence of the electronic system on the
behavior of the phonons cause that the basic symmetry relations
between the elastic constants (Cauchy relations) are no longer valid
\cite{Bro} and, therefore, even the basic results following from
the continuous medium approximation of Ref. \onlinecite{JMMM1} may be
questioned. All these factors cause that the simple Bloch approach
have to be used with a great caution in attempts of the {\em
quantitative} description of the electrical and thermal transport in
the alkali metals.

The most early investigations \cite{Dug60,Dug62} of the alkali metals
indicated that one can fix such parameters as those of Appendix B to
fit the experimental electrical resistivity to the Bloch--Gr\"uneisen
function, provided that the temperature is not too low. Thorough
analysis \cite{Dar,Gre} based on the measured phonon
spectra of the alkali metals \cite{Woo,Cow} showed later that the
problem is more complicated and the effects mentioned by us above
have to be in fact taken into account.  Therefore, even in case of
these simplest metals the correct quantitative theory based on first
principles is in fact impossible and a quantitative analysis have to
be based on a cross section for the scattering of the conduction
electrons written in terms of the phonon spectrum resulting from
neutron scattering data \cite{Woo,Cow}.  Precise experimental
measurements of the transport coefficients of the alkali metals at low
temperatures \cite{Vuc} support this conjuncture and reveals other
physical effects which should be taken into account for a proper
quantitative analysis in this temperature region.

As concerns the thermoelectric power our results indicate that other
factors than those in case of  the electrical resistivity are
important. The impurity scattering occurred less essential than
expected. Within the assumed model this is the parameter $\Es$ which
besides $\EF$ is solely responsible for the strength of the inelastic
scattering of electrons by phonons and controls the magnitude and
sign of $S$. The obtained temperature dependence of $S$ is the same
in character as that obtained experimentally \cite{Bar} for all the
the alkali metals except Li (which exhibits a peculiar behavior
\cite{Bar} but the Fermi surface of which comes closest to the
Brillouin zone boundaries). In search for a quantitative theory of
$S(T)$ of a specific metal one should also estimate the order of the
contribution of the phonon drag \cite{Bar}, that following from the
electron--phonon renormalization \cite{Kai} and the
electron--phonon--impurity interference. The described above factors
which influence the electrical resistivity in the alkali metals may
occur of secondary importance.

The effects of the impurity scattering manifest in most pronounced
way in the behavior of the thermal conductivity but due to the
validity the Wiedemann--Franz law in the limit of low temperatures
the effects of the impurity scattering can be easily extracted out of
our results at low temperatures for a comparison with experimental
ones \cite{Vuc}. As concerns the comparison of the values of
appropriate coefficients following from the theory and experimental
data for the alkali metals the same problems as in case of the
studies of the electrical resistivity may arise. The Umklapp process
are expected to be less dominant at low temperatures than in case of
the electrical resistivity but the electron--electron interactions
may occur essential for the thermal conductivity since both the
experimental and theoretical investigations indicate \cite{Vuc} that
electron--electron scattering in the alkali metals impedes
heat flow much more effectively than charge flow.


\begin{center} {\bf ACKNOWLEDGEMENTS} \end{center}

KD thanks FNRS and KBN for financial aid during part of this work.
MA thanks the Wroclaw PAN staff for its always warm welcome during 
various stays at the Institute
for Low Temperature and Structure Research.
Thanks also to various European programs, allowing for international exchanges.

\newpage
\begin{center}
{\bf APPENDIX A}
\end{center}

The transport coefficients considered in the paper follow from the
Onsager expressions for the electrical ${\rm\bf J}$ and thermal
${\rm\bf U}$ currents:

\ba \label{(A.1a)}
{\rm\bf J} = \sigma\,{\rm\bf E}+L\,\nabla T\:,\nonumber\\
{\rm\bf U} = \!M\,{\rm\bf E}+N\,\nabla T\:,
\ea
which in the ''experimental'' form are following

\ba \label{(A.1b)}
{\rm\bf E} = \rho\,{\rm\bf J}+S\,\nabla T\:,\nonumber\\
{\rm\bf U} = {\mit\pi}\,{\rm\bf J}-\kappa\,\nabla T\:
\ea
with $\mit\pi = - ST$ and

\ba\label{(A.2)}
\rho=1/ \sigma\:,\qquad
S=-\rho L\:,\qquad \kappa=M\rho L-N\:.
\ea

Within the Ziman variational method the coefficients of (\ref{(A.1a)}) are

\ba \label{(A.3)}
\sigma=\sum_{i,j=1}\,J_iQ_{ij}J_j\:, \mbox{\qquad\qquad}
L=-\frac1{T}\sum_{i,j=1}\,J_iQ_{ij}U_j\:, \nonumber\\
M=\sum_{i,j=1}\,U_iQ_{ij}J_j\:, &\mbox{\qquad\qquad}&
N=-\frac1{T}\sum_{i,j=1}\,U_iQ_{ij}U_j\:,
\ea
where $J_i$ and $U_i$ ($i,j = 1,2,\cdots$) are respectively the
electron and thermal trial currents defined by (\ref{11a})-(\ref{11b})
and $Q_{ij}$ is the inverse matrix to the scattering matrix $P_{ij}$,
i.e. $Q_{ij} = (P^{-1})_{ij}$.  In the approximation consisting in
abbreviating the combinations in (\ref{(A.3)}) to two components ($i,j
= 1,2$) we obtain (7)--(9) from (\ref{(A.2)}). For a degenerate
electron gas, i.e. for sufficiently large $z \cong \EF/ {\kB T}$, the
final expressions for the transport coefficients are (14) -- (16).
Notice that the thermal resistivity in particular is proportional to
$P_{22}$ .
\newpage
\begin{center}
{\bf APPENDIX B}
\end{center}

The functions describing the trial currents, as written below (15)-(16), are

\ba \label{B.1}
L_0(z)&=&(3/2)\,F_{1/2}(z)\:,\nonumber\\
L_1(z)&=& -(3/2)z\,F_{1/2}(z)+(5/2)\,F_{3/2}(z)\:,\\
L_2(z)&=&(3/2)z^2\,F_{1/2}(z)-5z\,F_{3/2}(z)+(7/2)\,F_{5/2}(z)\:.\nonumber
\ea
For completeness we also write down their asymptotic forms

\be \label{B.2}
L_0(z) \cong z^{3/2}\,,\mbox{\qquad}
L_1(z) \cong (\pi^2/2)\,z^{1/2}\,, \mbox{\qquad}
L_2(z) \cong (\pi^2/3)\,z^{3/2}\:.
\ee
The coefficients of the trial currents are $J_0$ = e $U_0$ with

\ba \label{A.3}
  U_0&=&\cases{%
\frac{1}{6 \pi^2} (\frac{2m}{\hbar^2})^{5/2}(\frac{\hbar}{m})^2\,\,\,
& $\qquad\qquad \,\,\,(\bv\cdot\boa)$ \cr
     \mbox{\hspace{3em}}& \mbox{\rm when the factor is}\,\,\,\cr
      \frac{1}{6\pi^2} 
(\frac{2m}{\hbar^2})^{5/2}(\frac{\hbar}{m})\,\,\, & $ \qquad\qquad
\, \, \,(\bk\cdot\boa)\:$
}
\ea
i.e. the upper case corresponding to the factor $(\bv\cdot\boa)$, and
the lower one to $(\bk\cdot\boa)$ in the trial function (\ref{6}).

The coefficient which determines the dimension and the order of the
magnitude of the phonon scattering matrix depends on the effective
electron mass, $m$, the volume of the primitive cell, $V$, its total
mass, $M$, the sound velocity $v_s$, the phonon wave vector cutoff,
$q_D$, and the deformation potential constant, $E_1$. It reads

\ba \label{A.4}
P_0^{(ph)}&=&\cases{  \frac{VE_1^2}{48 \pi^3 Mv_s}
(\frac{2m}{\hbar^2})^{2}(\frac{\hbar}{m})^2 q_D^5\,\,\, & $\qquad\qquad
(\bv\cdot\boa)\,\,\, $ \cr \mbox{\hspace{3em}}& \mbox{\rm when the factor is
}\,\,\, \cr \frac{VE_1^2}{48 \pi^3 Mv_s} (\frac{2m}{\hbar^2})^{2}
q_D^5\,\,\,  &
$ \qquad\qquad \,\,\, (\bk\cdot\boa) \:$. }
\ea

The coefficient $\rho_{0,ph} = P_0^{(ph)}/(e^2 U_0^2 \EF^3)$ in
(\ref{26}) can be represented by

  \be \label{A.5}
\rho_{0,ph} = 48 \pi (\mu)^2 (\frac{m_0}{\hbar^2 e^2})(\frac{m_0}{M})
VE_1^2 \tilde\rho_{0,ph}
\ee
where $m_0/(\hbar^2 e^2) = 9.113 \cdot 10^3$[eV$^{-2}\,$nm$^{-3}$],

\be \label{A.6}
\tilde\rho_{0,ph} =
(\frac {q_D}{2\kF})^6\, (v_s q_D)^{-1} = (\frac {q_D}{2\kF})^6\,\hbar
(\kB T_D)^{-1},
\ee
where $m_0$ is the free electron mass and $\mu = m/m_0$. For the value
of $T_D =200 K$ assumed for our representative data we have $\hbar (\kB
T_D)^{-1} = 3.437 \cdot 10^{-2}$\,$\Omega$ cm. In Table I we show some
representative data of $\tilde\rho_{0,ph}$ for $\EF= 1.5$\,eV, $T_D =
200$\,K, $\mu =1$ and the values of $\vep_s$ given in the table. They
correspond to those in Ref. \onlinecite{PRB96}, where numerical
relations between $\vep_s$ and other constants describing the phonon
system are presented.

The screened Coulomb potential scattering contribution to the transport is
described in terms of the dimensionless functions $\mit\Phi$ and $\mit\Psi$
defined by (\ref{4}) and (\ref{31}). The coefficient $P_0^{(imp)}$ of the
scattering matrix elements is

\ba \label{A.7}
P_0^{(imp)}&=&\cases{  \frac{1}{3
\pi \hbar}\,n_{imp}Z^2 e^4 (\frac{2m}{\hbar^2})^{2}(\frac{\hbar}{m})^2\,\,\, &
$\qquad\qquad (\bv\cdot\boa)\,\,\, $ \cr \mbox{\hspace{3em}}&
\mbox{\rm when the
factor is}\,\,\, \cr \frac{1}{3 \pi \hbar}\,n_{imp}Z^2 e^4
(\frac{2m}{\hbar^2})^{2}\,\,\,  & $ \qquad\qquad \,\,\,
(\bk\cdot\boa) \:$. }
\ea
where $n_{imp}$ is the impurity concentration (their number per unit
volume). The residual resistivity $\rho^{(imp)} = P_{11}^{(imp)}/ (e^2
U_0^2 \EF^3)$ can be represented by

\be \label{A.8}
\rho^{(imp)} = \frac{Z^2 n_{imp}}{n_e\mu}\,\tilde\rho_{imp}\,
\mit\Phi(2\kF /\lambda)
\ee
where $n_e$ is the number of the conduction electrons per unit volume, with

\be \label{A.9}
\tilde\rho_{imp} = \frac{12 \pi^2 e^2}{m_0 {v_F}^3}\,;
\ee

$v_F$ is the Fermi velocity, and $2
\kF/\lambda \cong 1$ as we mentioned in Sec. II. For the data 
represented in Table I (free
electron gas, $\EF = 1.5$\,eV) we have $\tilde\rho_{imp} =
0.075\,\Omega$~cm.

In order to relate the magnitude of $\mit\Gamma$ in (25)--(26) with
the number of impurities in the sample we represent it as $\tilde\varphi$

\be
\mit\Gamma = \frac{\rho^{(imp)}}{\rho_{0,ph}} =
\frac{\tilde\rho_{imp}}{\rho_{0,ph}}\,\mit\Phi(2\kF/\lambda)\,\frac{Z^2
n_{imp}}{n_e\mu} \equiv \tilde\varphi\, \mit\Phi(2\kF/\lambda)\,\frac{Z^2
n_{imp}}{n_e\mu}
\ee
and shall estimate $\tilde\varphi$. It means that $\varphi$ (31) is
$\varphi = (Z^2 n_{imp}/n_e \mu) \tilde\varphi$. For the mass of the
unit cell $M=10^6 m_0$, its volume $V= 0.1$\,nm$^3$ and $E_1=(2/3)\EF
= 1$\,eV we have $\rho_{0,ph} \cong 0.1\,\tilde\rho_{0,ph}$. This is
an underestimated value in many cases, thus we may assume $\rho_{0,ph}
= a\,\tilde\rho_{0,ph}$ with $0.1 < a < 1$. For $\EF= 1.5$\,eV, $T_D =
200$\,K $\mu =1$ the value of $\tilde\rho_{imp}$ is: $\tilde\rho_{imp}
\cong 0.1\,\Omega$cm.  The latter value along with those ones for
$\rho_{0,ph}$ following from the values of $\tilde\rho_{0,ph}$
presented the Table I yields an estimate of $\tilde\varphi$.

Finally, for information as to the magnitude of the thermal conductivity
represented in dimensionless units in the figures we give the magnitude of

\be
\tilde\kappa_{0,ph} = \frac{\pi^2}{3}\,\frac{{\cal L}_0
T_D}{\tilde\rho_{0,ph}}
\ee
for the Debye temperature $T_D = 200$\,K in the last column of Table
I.

\newpage
\begin{table}
\caption[A] {\small The values of the parameters allowing to estimate the
magnitude of the electrical resistivity and electronic thermal
conductivity for a
free electron effective mass, when the Fermi energy is $\EF = 1.5$~eV and the
Debye temperature $T_D = 200$\,K.} \vspace{2ex} \begin{tabular}{cccccc} \hline
&&&&&\\[-2ex] $no.$ & $\vep_s$\,[K] & $\tilde\rho_{0,ph}$\,[$\Omega$ cm] &
$\tilde\varphi$ & $\tilde\kappa_{0,ph}$\,[W/mK] \\[1ex] \hline
&&&&&\\[-2ex] (i)
& 1.50 & 1.93 $\cdot 10^{-3}$ & $ 10^{0}$ -- $10^{1}$ & 2.54 $\cdot
10^{-5}$ &\\
(ii) & 3.00 & 2.39 $\cdot 10^{-4}$& $ 10^{1}$ -- $10^{2}$ & 2.04
$\cdot 10^{-4}$
& \\ (iii) & 6.00 & 2.99 $\cdot 10^{-5}$&  $ 10^{2}$ -- $10^{3}$ & 1.64 $\cdot
10^{-3}$ &\\ (iv) & 12.00 & 3.79 $\cdot 10^{-6}$&  $ 10^{3}$ -- $10^{4}$ & 1.23
$\cdot 10^{-2}$ &\\ (v) & 24.0 & 4.58 $\cdot 10^{-7}$& $ 10^{4}$ -- $10^{5}$ &
1.07 $\cdot 10^{-1}$ &\\ (vi) & 48.0 &  5.76 $\cdot 10^{-8}$& $10^{5}$ --
$10^{6}$ & 8.48 $\cdot 10^{-1}$ & \\[1ex] \end{tabular}
\end{table}

\newpage
\noindent
{\bf Figure Captions}

\vspace*{0.6cm}
\noindent
{\bf Figure 1} --- Temperature
dependence of $\rho/\rho_{0,ph}$ resulting from Eq.(26) for values of the
parameters $2\kF/\lambda = 0.75$ and the relative scattering strength
$\varphi =
0.0, 0.2, 0.5, 1.0$. They correspond to the values of $\mit\Gamma =
\varphi\mit\Phi$(0.75) which are 0.0, 0.0173, 0.0431, 0.0862, respectively.

\vspace*{0.6cm}
\noindent
{\bf Figure 2} ---   Temperature
dependence of $S/S_{k,e}$ for values of the parameters as in Fig.1
with $\EF=1.5$
eV, $\Es=6$ K, and the effective electron mass equal to that of a
free electron.
$S_{k,e}$ = ${k_B}/e$ = 86 $\mu$ V/K.

\vspace*{0.6cm}

\noindent
{\bf Figure 3} --- Temperature
dependence of the reduced thermal conductivity below $T_D$, for $2\kF/\lambda =
0.75$ and $\varphi = 0.0,\, 0.2,\, 0.5,\, 1.0,\, 2.0$; the corresponding values
of $\mit\Gamma$ are 0.0,\, 0.0173,\, 0.0431,\, 0.0862,\, 0.173,
respectively. For
$\mit\Gamma$ = 0,  the value of $\kappa$ is infinite. Note that for
the magnitude
of $\mit\Gamma$ as large as 2.0 there is a plateau of $\kappa$ instead of the
maximum at $T \cong 0.1 T_D$.

\vspace*{0.6cm}

\noindent
{\bf Figure 4} --- Temperature
dependence of the reduced thermal conductivity in the same temperature range as
that displayed for the thermoelectric power in Fig.2. The value of
$2\kF/\lambda$
is the same as in Fig.3, i.e. 0.75, and $\varphi = 0.0,\, 0.2,\, 0.5,\, 1.0,\,
2.0$.

\vspace*{0.6cm}

\noindent
{\bf Figure 5} --- Temperature dependence of the
reduced Lorenz number in the same temperature range as that of the
thermoelectric
power in Fig.2. The value of $2\kF/\lambda$ is 0.75 as in Fig.3 , and
$\varphi =
0.0,\, 0.2,\, 0.5,\, 1.0,\, 2.0$.

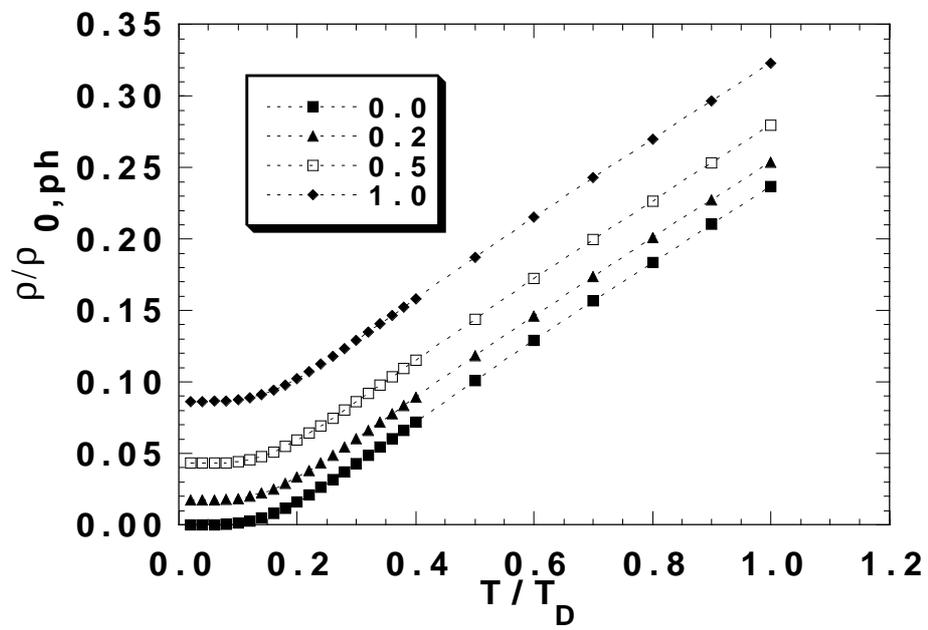
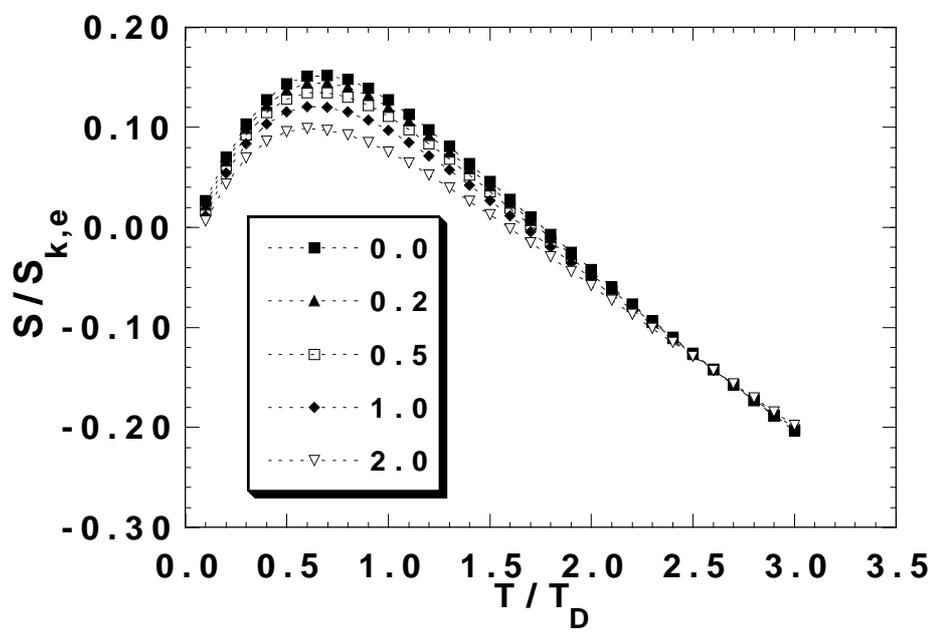
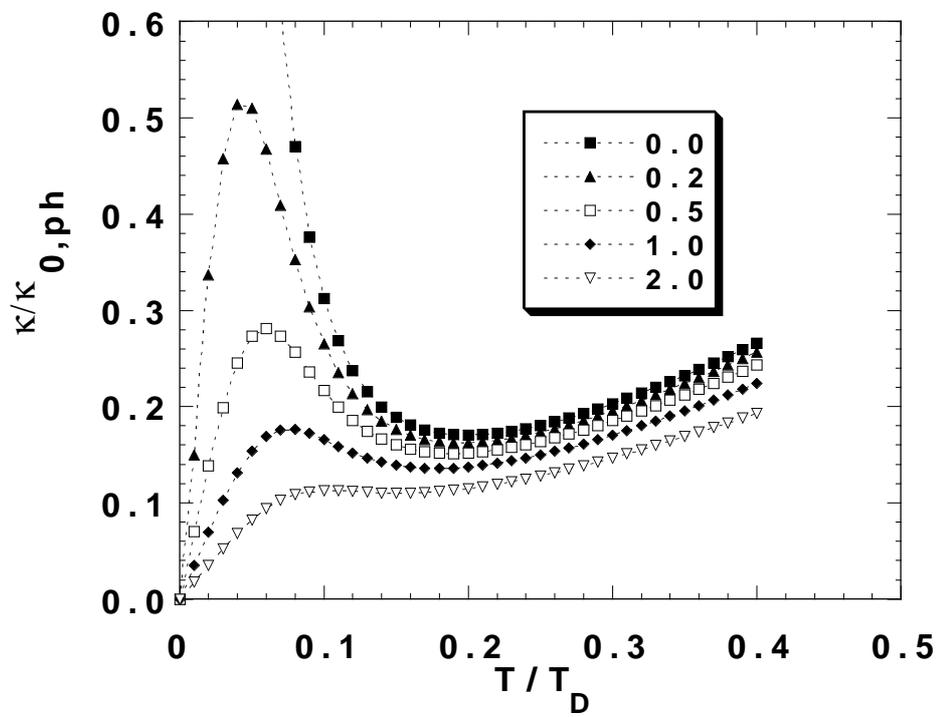
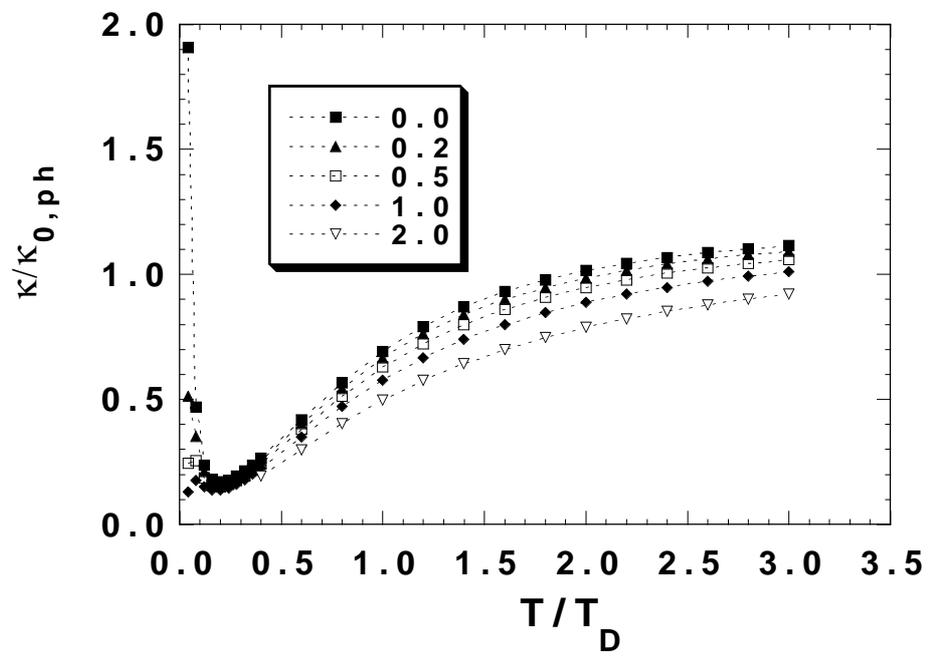
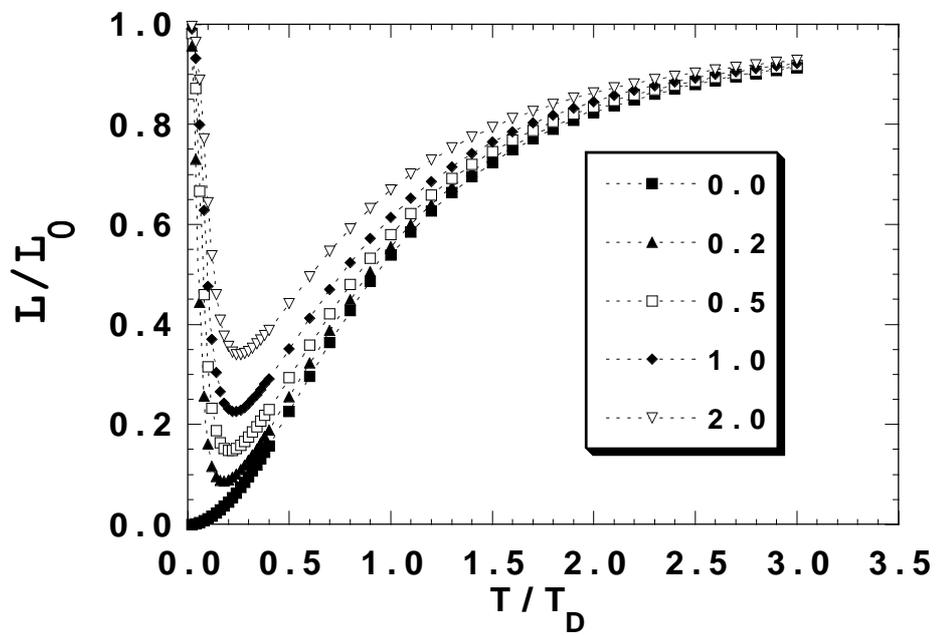
\end{document}